\documentclass[aps,prb,a4paper,preprint,showpacs]{revtex4}
\usepackage{amsmath}
\usepackage{amsfonts}
\usepackage{amssymb}
\usepackage{slashbox}
\usepackage{graphicx}
\usepackage{amsbsy}

\begin{document}
\title{AC square-wave, sawtooth-wave, and triangle-wave fields driven adiabatic quantum pumps in nanowire structures}
\author{Rui Zhu\renewcommand{\thefootnote}{*} \footnote{Corresponding author. Electronic address:
rzhu@scut.edu.cn}, Xiao-Kang Zhang,  and Xiaowen Chen  }
\address{Department of Physics, South China University of Technology,
Guangzhou 510641, People's Republic of China }

\begin{abstract}

A dc current can be pumped through a nanostructure by two
out-of-phase ac driving fields. In this approach, we explore the
pumped current properties driven by square-wave, sawtooth-wave, and
triangle-wave ac fields in nanowire structures by Fourier expansion.
The theory can be applied to arbitrary driving fields. It is found
that the pumped current varies with the phase difference as
square-wave and sawtooth-wave functions, respectively, under
corresponding driven fields. The sinusoidal relation governs in the
pump driven by triangle-wave oscillation. Devices based on
square-wave driven quantum pumps may have potential applications in
digital information at nanoscale dimensions.

\end{abstract}


\maketitle

\narrowtext

\section{Introduction}

Quantum pumping is a transport mechanism which induces dc charge and
spin currents in a nano-scale conductor in the absence of a bias
voltage by means of a time-dependent control of some system
parameters. Research on quantum pumping has attracted heated
interest since its experimental realization in an open quantum
dot\cite{Ref2, Ref3, Ref4, Ref5, Ref6, Ref7, Ref8, Ref9, Ref10,
Ref11, Ref12, Ref13, Ref14, Ref15, Ref16, Ref17, Ref18, Ref19,
Ref20, Ref21, Ref22, Ref23, Ref24, Ref25, Ref26, Ref27, Ref28,
Ref29, Ref31, Ref33, Ref36, Ref37}. Harmonic-force-driven pumping
properties in various nano-scale structures were investigated such
as the magnetic-barrier-modulated two dimensional electron
gas\cite{Ref5}, mesoscopic one-dimensional wire\cite{Ref7, Ref23},
quantum-dot structures\cite{Ref6, Ref12, Ref13, Ref29, Ref30,
Ref37}, mesoscopic rings with Aharonov-Casher and Aharonov-Bohm
effect\cite{Ref8}, magnetic tunnel junctions\cite{Ref11}, chains of
tunnel-coupled metallic islands\cite{Ref26}, the nanoscale helical
wire\cite{Ref27},the Tomonaga-Luttinger liquid\cite{Ref25}, and
garphene-based devices\cite{Ref21, Ref22}. Recently, an ac-dc
voltage probe to quantum driven systems was demonstrated and the
local current profile was predicted\cite{Ref39}. Theory and
experiments demonstrate that the harmonically-driven dc current
produced at zero bias is sinusoidal in the phase difference between
the two ac voltages in the adiabatic and weak pumping regime.

Recently, symmetries and transport properties of quantum pumps with
biharmonic driving were investigated by several authors\cite{Ref36}.
Keldysh non-equilibrium Green's function and the scattering-matrix
method can be used to derive the pumped current in certain cases.
However, quantum pumping driven by arbitrary fields is less
investigated and that driven by fields with sharp fall and increase
such as the square wave and sawtooth wave is beyond former approach.
The square wave, sawtooth wave, and triangle wave are three typical
non-sinusoidal waveforms. The square wave is widely used in
electronics and signal processing, such as timing references or
``clock signals". The sawtooth wave is one of the best waveforms to
use for synthesizing musical sounds. Their significance in signal
processing prompted our exploration of the quantum-pumping
properties driven by two out-of-phase square-wave, sawtooth-wave,
and triangle-wave ac fields and seeking the potential applications.
Relevant theory can be extended to arbitrary-field-driven quantum
pumps.

\section{Theoretical formulation}

The square wave is a kind of non-sinusoidal waveform. Using Fourier
expansion we can write an ideal square wave as an infinite series of
the form
\begin{equation}
x_{Square} \left(\omega t \right) = \sum\limits_{k = 1}^\infty
{\frac{{\sin \left[ {\left( {2k - 1} \right)\omega t}
\right]}}{{\left( {2k - 1} \right)}}} ,
\end{equation}
with the angular frequency of $\omega$ and the amplitude of $\pi /4$
(see the inset of Fig. 1). We consider the response of a mesoscopic
phase-coherent sample to two slowly oscillating square-wave external
fields $X_{j}(t)$ (gate potential, magnetic flux, etc.)\cite{Ref4},
which can be expressed as
\begin{equation}
\begin{array}{l}
 X_j \left(  t \right) = X_{0,j}  + \frac{{X_{\omega ,j} }}{{2i}}\sum\limits_{k = 1}^\infty  {\frac{{\exp \left[ {i\left( {2k - 1} \right)\left( {\omega t - \varphi _j } \right)} \right]}}{{\left( {2k - 1} \right)}}}  \\
  \hspace{1.2 cm} - \frac{{X_{\omega ,j} }}{{2i}}\sum\limits_{k = 1}^\infty  {\frac{{\exp \left[ { - i\left( {2k - 1} \right)\left( {\omega t - \varphi _j } \right)} \right]}}{{\left( {2k - 1} \right)}}} , \\
 j = 1,2. \\
 \end{array}
\end{equation}
$X_{0,j}$ and $X_{\omega ,j}$ measure the static magnitude and the
square-wave ac driving amplitude of the two parameters,
respectively. The phase difference between the two drivers is
defined as $\phi  = \varphi _1 - \varphi _2 $. The mesoscopic
conductor is connected to two reservoirs at zero bias. The
scattering matrix $\hat s$ being a function of parameters $X_{j}(t)$
depends on time.

We suppose an adiabatic quantum pump, i.e., the external parameter
changes so slowly that up to corrections of order $\hbar \omega /
\gamma$ ($\gamma$ measures the escape rate), we can apply an instant
scattering description using the scattering matrix $\hat s\left( t
\right)$ frozen at some time $t$. And we assume that the amplitude
${X_{\omega ,j} }$ is small enough to keep only the terms linear in
${X_{\omega ,j} }$ in an expansion of the scattering
matrix\cite{Ref4}:
\begin{equation}
\hat s\left( t \right) \approx \hat s\left( {X_{0,j} } \right) +
\sum\limits_{k = 1}^\infty  {\hat s_{ - k\omega } e^{i\left( {2k -
1} \right)\omega t} }  - \sum\limits_{k = 1}^\infty  {\hat s_{ +
k\omega } e^{ - i\left( {2k - 1} \right)\omega t} } .
\end{equation}
In the limit of small frequencies the amplitudes $\hat s_{ \pm
k\omega }$ can be expressed in terms of parametric derivatives of
the on-shell scattering matrix $\hat s$,
\begin{equation}
\hat s_{ \pm k\omega }  = \sum\limits_{j = 1,2} {\frac{{X_{\omega
,j} }}{{2i\left( {2k - 1} \right)}}e^{ \pm i\left( {2k - 1}
\right)\varphi _j } \frac{{\partial \hat s}}{{\partial X_j }}} .
\end{equation}
The expansion, Eq. (3), includes sideband formation up to infinite
orders, which implies that a scattered electron can absorb numerous
energy quanta of $\hbar \omega$ before it leaves the scattering
region. We would like to point out that square waves contain a wide
range of harmonics. Therefore the scattering matrix naturally
includes high-order Fourier terms even in the linear-response limit.
The conduction bandwidth of a semiconductor nanowire is of the order
of meV within the non-dissipation regime\cite{Ref34} and an $\hbar
\omega$ is of the order of $10^{-6}$ meV for an MHz frequency. The
square-wave shape achieves acceptable approximation with 100 orders
of harmonics present. The setting time is extremely small, while the
Gibbs phenomenon is unavoidable. Thus formed sideband of even $100
\hbar \omega$ constitutes a slight broadening in the conductance and
the absorption/emission of a $100 \hbar \omega$ quantum is
physically justifiable. The effect of higher-order harmonics can be
neglected.

The pumped current depends on the values of the scattering matrix
within the energy interval of the order of ${\rm{max}}(k_B T, 100
\hbar \omega)$ near the Fermi energy. In the low-temperature limit
($T \to 0$) and low-frequency limit ($\omega \to 0$), the scattering
matrix can be assumed to be energy independent within a $100\hbar
\omega  \sim 10^{ - 4} {\rm{meV}}$ deviation as a driving frequency
of the order of MHz is considered.

The mesoscopic scatterer is coupled to two reservoirs with the same
temperatures $T$ and electrochemical potentials $\mu$. Electrons
with the energy $E$ entering the scatterer are described by the
Fermi distribution function $f_{0} (E)$, which approximates a step
function at a low temperature. Due to the interaction with an
oscillating scatterer, an electron can absorb or emit energy quanta
that changes the distribution function. A single transverse channel
in one of the leads is considered. Applying the hypothesis of an
instant scattering, the scattering matrix connecting the incoming
and outgoing states can be written as
\begin{equation}
\hat b_\alpha  \left( t \right) = \sum\limits_\beta  {s_{\alpha
\beta } \left( t \right)\hat a_\beta  \left( t \right)}.
\end{equation}
Here $s_{\alpha \beta } $ is an element of the scattering matrix
$\hat s$; the time-dependent operator is $\hat a_\alpha  \left( t
\right) = \int {dE\hat a_\alpha  \left( E \right)e^{{{ - iEt}
\mathord{\left/
 {\vphantom {{ - iEt} \hbar }} \right.
 \kern-\nulldelimiterspace} \hbar }} } $,
and the energy-dependent operator ${\hat a_\alpha  \left( E
\right)}$ annihilates particles with total energy $E$ incident from
the $\alpha$ lead into the scatter and obey the following
anticommutation relations
\begin{equation}
\left[ {\hat a_\alpha ^\dag  \left( E \right),\hat a_\beta  \left(
{E'} \right)} \right] = \delta _{\alpha \beta } \delta \left( {E -
E'} \right).
\end{equation}
Note that above expressions correspond to single- (transverse)
channel leads and spinless electrons. For the case of many-channel
leads each lead index ($\alpha $, $\beta$, etc.) includes a
transverse channel index and any repeating lead index implies
implicitly a summation over all the transverse channels in the lead.
Similarly an electron spin can be taken into account.

Using Eqs. (3) to (5) and after a Fourier transformation we obtain
\begin{equation}
\begin{array}{l}
 \hat b_\alpha  \left( E \right) = \sum\limits_\beta  {\hat s\left( {X_{0,j} } \right)\hat a_\beta  \left( E \right)}  \\
 \hspace {1.3 cm} + \sum\limits_\beta  {\sum\limits_{k = 1}^\infty  {\hat s_{ - k\omega } \hat a_\beta  \left[ {E + \left( {2k - 1} \right)\hbar \omega } \right]} }  \\
  \hspace {1.3 cm} - \sum\limits_\beta  {\sum\limits_{k = 1}^\infty  {\hat s_{ + k\omega } \hat a_\beta  \left[ {E - \left( {2k - 1} \right)\hbar \omega } \right]} } . \\
 \end{array}
\end{equation}
The distribution function for electrons leaving the scatterer
through the lead $\alpha$ is $f_\alpha ^{\left( {out} \right)}
\left( E \right) = \left\langle {\hat b_\alpha ^\dag  \left( E
\right)\hat b_\alpha  \left( E \right)} \right\rangle $, where
$\left\langle  \cdots  \right\rangle $ means quantum-mechanical
averaging. Substituting Eq. (7) we find
\begin{equation}
\begin{array}{l}
 f_\alpha ^{\left( {out} \right)} \left( E \right) = \sum\limits_\beta  {\left| {\hat s\left( {X_{0,j} } \right)} \right|^2 f_0 \left( E \right)}  \\
  \hspace {1.8 cm} + \sum\limits_\beta  {\sum\limits_{k = 1}^\infty  {\left| {\hat s_{ - k\omega } } \right|^2 f_0 \left[ {E + \left( {2k - 1} \right)\hbar \omega } \right]} }  \\
  \hspace {1.8 cm} + \sum\limits_\beta  {\sum\limits_{k = 1}^\infty  {\left| {\hat s_{ + k\omega } } \right|^2 f_0 \left[ {E - \left( {2k - 1} \right)\hbar \omega } \right]} } . \\
 \end{array}
\end{equation}
 The distribution function for outgoing carriers is a
nonequilibrium distribution function generated by the nonstationary
scatterer. The Fourier amplitudes of the scattering matrix ${\left|
{\hat s_{ - k\omega ,\alpha \beta } } \right|^2 }$ (${\left| {\hat
s_{ + k\omega ,\alpha \beta } } \right|^2 }$) is the probability for
an electron entering the scatterer through the lead $\beta$ and
leaving the scatterer through the lead $\alpha$ to emit (to absorb)
an energy quantum of $ \hbar k \omega $. $\left| {\hat s\left(
{X_{0,j} } \right)} \right|^2$ is the probability for the same
scattering without the change of energy.

Using the distribution functions $f_{0} (E)$ for incoming electrons
and $f_{\alpha} ^{out} (E)$ for outgoing electrons, the pumped
current measured at lead $\alpha$ reads
\begin{equation}
I_p  = \frac{e}{{2\pi \hbar }}\int_0^\infty  {\left\langle {\hat
b_\alpha ^\dag  \left( E \right)\hat b_\alpha  \left( E \right)}
\right\rangle  - \left\langle {\hat a_\alpha ^\dag  \left( E
\right)\hat a_\alpha  \left( E \right)} \right\rangle dE}.
\end{equation}
Substituting Eqs. (8) and (4) into Eq. (9) we get
\begin{equation}
\begin{array}{l}
 I_p  = \frac{{ie\omega }}{{4\pi }}\sum\limits_\beta  {\sum\limits_{k = 1}^\infty  {\sum\limits_{j_1 ,j_2 } {\frac{{X_{\omega ,j_1 } X_{\omega ,j_2 } }}{{\left( {2k - 1} \right)}}} \sin \left[ {\left( {2k - 1} \right)\left( {\varphi _{j_1 }  - \varphi _{j_2 } } \right)} \right]\frac{{\partial \hat s_{\alpha \beta }^* }}{{\partial X_{j_1 } }}\frac{{\partial \hat s_{\alpha \beta } }}{{\partial X_{j_2 } }}} }  \\
 \hspace{0.5 cm} = \frac{{ie\omega }}{{4\pi }}\sum\limits_\beta  {\sum\limits_{j_1 ,j_2 } {X_{\omega ,j_1 } X_{\omega ,j_2 } } \frac{{\partial \hat s_{\alpha \beta }^* }}{{\partial X_{j_1 } }}\frac{{\partial \hat s_{\alpha \beta } }}{{\partial X_{j_2 } }}x_{Square} \left( {\varphi _{j_1 }  - \varphi _{j_2 } } \right).}  \\
 \end{array}
\end{equation}
Quantum pumping properties driven by ac square-wave fields are
demonstrated in Eq. (10). It can be directly seen that the pumped
current varies with the phase difference of the two drivers in a
square-wave pattern. The magnitude of the pumped current is
modulated by the driving amplitude and the scattering matrix
derivatives. The wave function of the pumped current is exactly
square-shaped with all harmonics involved. Physically, in the
Fourier series, energy quanta of $ \hbar k \omega $ emission
(absorption) processes with $k$ ranging many orders determine the
pumped current. High-order Fourier terms diminish due to a large
denominator produced in the Fourier expansion. Numerical results
will show that expanding to the 100th harmonic generates a
square-wave shape with extremely small ringing effect and
higher-order harmonics can be neglected. The relation between the
pumped current and the ac driving amplitude $X_{\omega ,j} $ is
linear in the adiabatic weak-modulation limit. The linear dependence
of the pumped current on the oscillation frequency holds valid as
the adiabatic approximation is considered.

Following the same theory and similar derivation, the pumped current
driven by two out-of-phase sawtooth-wave and triangle-wave fields
can be obtained as in Eqs. (11) and (12), respectively.
\begin{equation}
\begin{array}{l}
 I_p  = \frac{{ie\omega }}{{4\pi }}\sum\limits_\beta  {\sum\limits_{k = 1}^\infty  {\sum\limits_{j_1 ,j_2 } {\frac{{X_{\omega ,j_1 } X_{\omega ,j_2 } }}{k}} } }   \sin \left[ {k\left( {\varphi _{j_1 }  - \varphi _{j_2 } } \right)} \right]\frac{{\partial \hat s_{\alpha \beta }^* }}{{\partial X_{j_1 } }}\frac{{\partial \hat s_{\alpha \beta } }}{{\partial X_{j_2 } }} \\
 \hspace{0.5 cm} = \frac{{ie\omega }}{{4\pi }}\sum\limits_\beta  {\sum\limits_{j_1 ,j_2 } {X_{\omega ,j_1 } X_{\omega ,j_2 } \frac{{\partial \hat s_{\alpha \beta }^* }}{{\partial X_{j_1 } }}\frac{{\partial \hat s_{\alpha \beta } }}{{\partial X_{j_2 } }}} x_{Sawtooth} \left( {\varphi _{j_1 }  - \varphi _{j_2 } } \right)} , \\
 \end{array}
\end{equation}
\begin{equation}
I_p  = \frac{{ie\omega }}{{4\pi }}\sum\limits_\beta  {\sum\limits_{k
= 1}^\infty  {\sum\limits_{j_1 ,j_2 } {\frac{{X_{\omega ,j_1 }
X_{\omega ,j_2 } }}{{\left( {2k - 1} \right)^3 }}} } }   \sin \left[
{\left( {2k - 1} \right)\left( {\varphi _{j_1 }  - \varphi _{j_2 } }
\right)} \right]\frac{{\partial \hat s_{\alpha \beta }^*
}}{{\partial X_{j_1 } }}\frac{{\partial \hat s_{\alpha \beta }
}}{{\partial X_{j_2 } }}.
\end{equation}
In Eq. (11), $x_{Sawtooth}$ is the function of a sawtooth wave
analogous to Eq. (1). It can be seen from Eqs. (10) and (11) that in
our approach the Fourier expansion in the signal function can be
converted back to the original wave form as a function of the phase
difference in the pumped current for square-wave and sawtooth-wave
signals. For triangle-wave signals, the higher harmonics roll off
much faster than in square and sawtooth waves (proportional to the
inverse square of the harmonic number as opposed to just the
inverse). In pumping mechanisms, that attenuation is magnified with
each harmonic term in the pumped current proportional to the inverse
cubic of the harmonic number and original triangle-wave shape can
not be reproduced.

It is worth mentioning that former scattering approach\cite{Ref3} to
quantum pumping gives a general formula of the pumped current for
arbitrary parameter variation pattern. However, for square- and
sawtooth-wave modulation, sharp fall and increase presents in the
parameter variation. Without Fourier series expansion, direct
parameter derivatives diverge. If we use the Fourier series,
numerical solution of the area integral in the parameter space
becomes extremely difficult\cite{Ref38}. In the Keldysh
formalism\cite{Ref36, Ref37}, the Green's function is difficult to
be derived from the equation of motion with the square-wave
time-dependent potential modulating the Hamiltonian. The problem is
numerically solvable by our approach and the numerical efficiency is
markedly improved .

In the next section, numerical results of the pumped current in a
two-oscillating-potential-barrier modulated nanowire are presented.

\section{Numerical results and interpretations}

We consider a nanowire modulated by two gate potential barriers with
equal width $L=20$ {\AA} separated by a $2L=40$ {\AA} width well.
The electrochemical potential of the two reservoirs $\mu$ is set to
be $60$ meV according to the resonant level within the
double-barrier structure. The two oscillating parameters in Eq. (2)
correspond to the two ac driven potential gates. We set the static
magnitude of the two gate potentials $U_{0,1}  = U_{0,2} = U_0  =
100$ meV and the ac driving amplitude of the modulations equal with
$U_{\omega ,1}  = U_{\omega ,2} = U_\omega $.

Figs. 1, 2, and 3 present numerical results of the dc current pumped
by square-wave, sawtooth-wave, and triangle-wave drivers,
respectively. Corresponding driving signals are shown in insets.
Numerical calculation takes into account the foremost 100 orders of
the Fourier series. As shown in Fig. 1, the pumped current varies
with the phase difference between the two drivers in a square-wave
pattern when the driving signal is a square wave, which is a direct
result of former analysis [see Eq. (10) and previous derivation].
Phase difference has a modulation to the time dependence of the
driving forces. The pumped current varies with the phase difference
in a sinusoidal pattern when the driving field is a sinusoidal wave
as a function of time. When higher-order harmonics are present, the
effect of the phase difference is similar to the sinusoidal harmonic
when their strength depends linearly-inversely on the order number.
This behavior is also demonstrated in the sawtooth wave case, as
shown in Fig. 2 and formulized in Eq. (11). Different from the
square-wave and sawtooth-wave fields, the pumped current driven by
triangle varies with the phase difference of the two ac gate
voltages in a sinusoidal pattern as the higher harmonics roll off
much faster [see Fig. 3 and Eq. (12)].

Our theory can be extended to arbitrary-field-driven quantum pumps.
When the strength of all orders of the Fourier expansion is
proportional to the inverse of the harmonic number, the signal
pattern as a variation of the time can be reproduced in the pumped
current as a variation of the phase difference. In other cases, the
reproduction would not occur. It can also be inferred from our
theory that the variation of the pumped current as a function of the
phase difference would deviate from the wave pattern even for
sinusoidal, square-wave, and sawtooth-wave ac fields when the
driving force is strong beyond the linear response
limit\cite{Ref33}.

In digital information technology, square waves and sawtooth waves
are two types of widely used digital signals. Classic oscilloscopes
and phase detectors are usually based on LC circuits. In our
approach, the characteristics of the pumped current bear information
of both the wave pattern and wave phase of the input ac signals.
Potential development of the currently smallest quantum
oscilloscopes and phase detectors can be foreseen from the
theoretical approach.

\section{Conclusions}

Two out-of phase harmonic ac fields driven quantum pumps have a
sinusoidal dependence of the pumped current on the phase difference
of the two drivers. We developed the scattering-matrix method to
explore the pumping properties driven by square-wave, sawtooth-wave,
and triangle-wave ac fields by harmonic expansion of the
time-dependent signals. It is demonstrated that the pumped current
varies with the phase difference of the two ac fields as a
square-wave and sawtooth-wave pattern driven by corresponding wave
signals. The theory can be extended to arbitrary field driven
quantum pumps, which infers that the signal time-dependent pattern
can be reproduced in the phase dependent pumped current in certain
cases. The wave pattern and phase information of the input signal
carried by the pumped current in the proposed configuration suggests
its potential applications.

\section{Acknowledgements}

This project was supported by the National Natural Science
Foundation of China (No. 11004063), the Fundamental Research Funds
for the Central Universities, SCUT (No. 2009ZM0299), the Nature
Science Foundation of SCUT (No. x2lxE5090410) and the Graduate
Course Construction Project of SCUT (No. yjzk2009001).

\clearpage

\begin{figure}[h]
\includegraphics{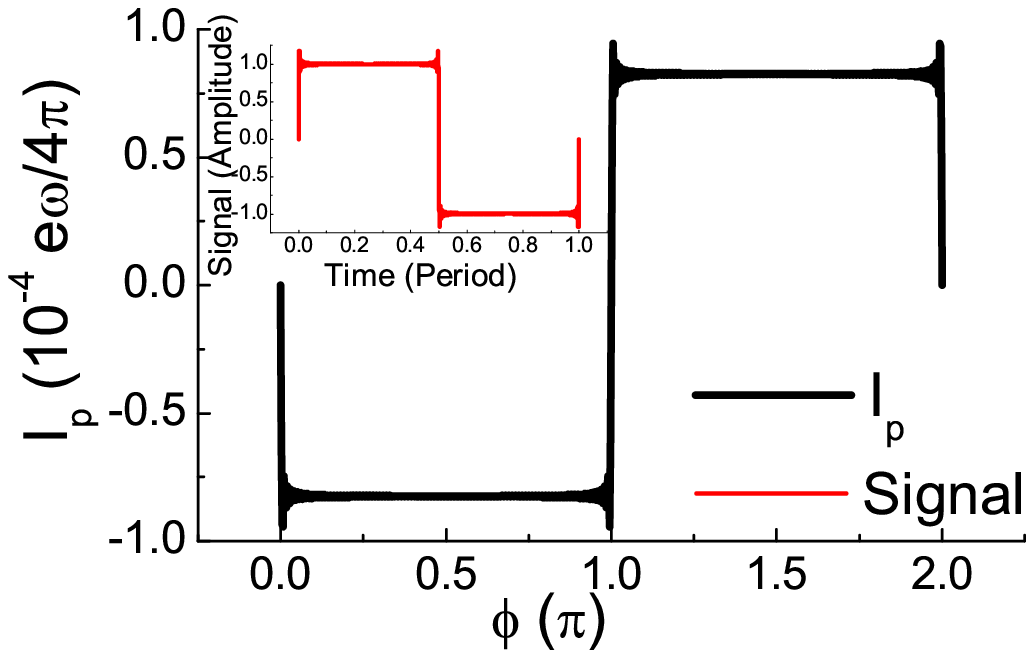}
\caption{Pumped current driven by ac square-wave fields as a
function of the phase difference
 between the two modulations. Inset is the time-variation of the wave signal.
  The foremost 100 harmonics are considered in numerical treatment.}
\end{figure}

\clearpage

\begin{figure}[h]
\includegraphics{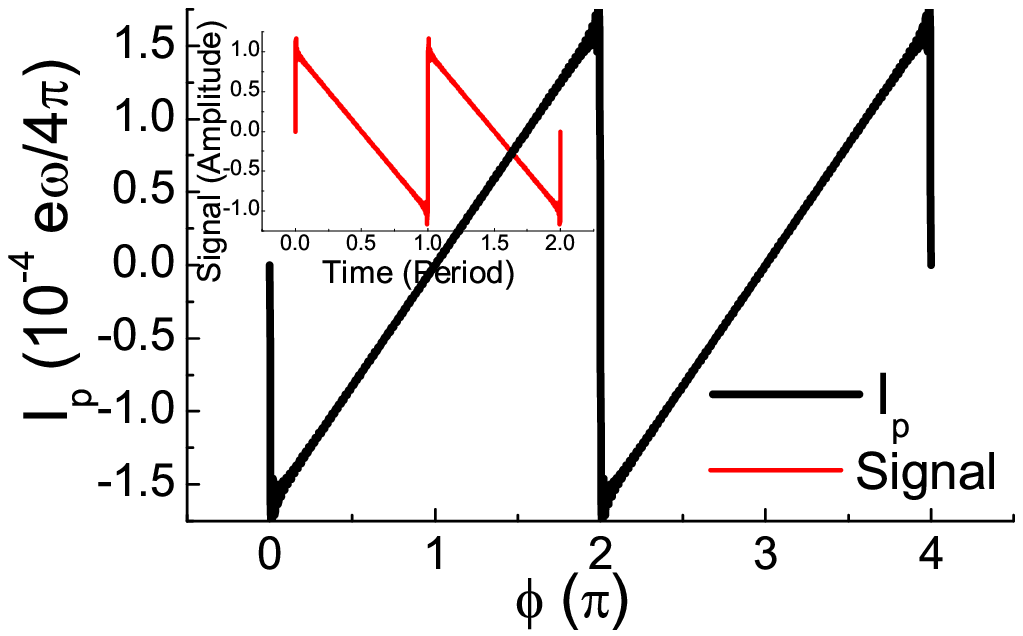}
 \caption{Pumped current driven by ac sawtooth-wave fields as a
function of the phase difference
 between the two modulations. Inset is the time-variation of the wave signal.
 The foremost 100 harmonics are considered in numerical treatment.
 }
\end{figure}

\begin{figure}[h]
\includegraphics{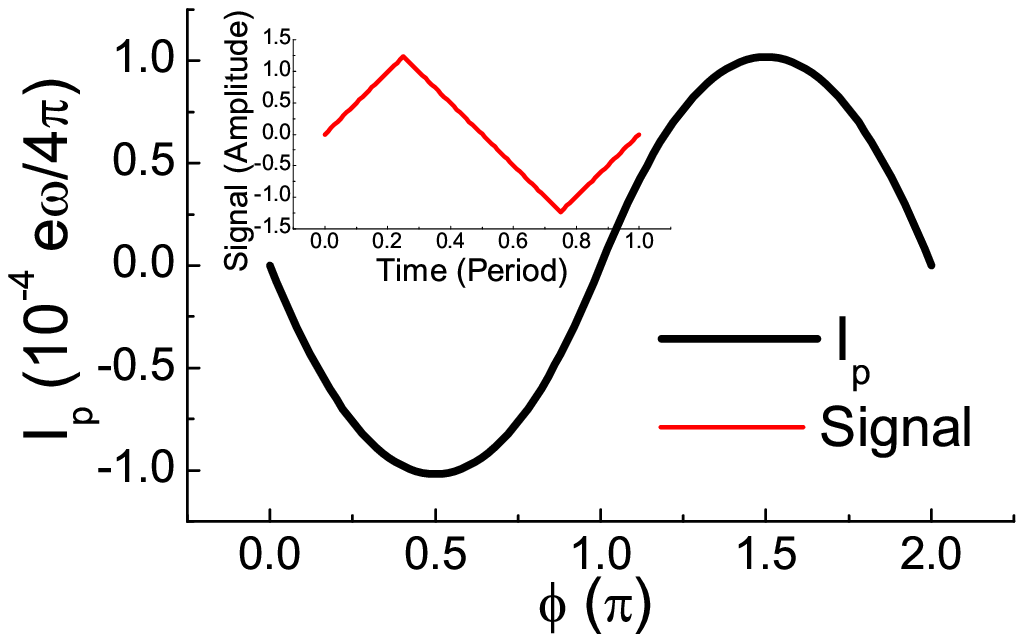}
 \caption{Pumped current driven by ac triangle-wave fields as a
function of the phase difference
 between the two modulations. Inset is the time-variation of the wave signal.
 The foremost 100 harmonics are considered in numerical treatment.}
\end{figure}


\begin{references}
\bibitem{Ref38} By Monte Carlo algorithms that numerical integration may
become viable if additional programming efforts are made.
\end{references}
\end{document}